\documentclass[prb,twocolumn,showpacs,floatfix,superscriptaddress]{revtex4}
\usepackage{epsfig}
\usepackage{amsmath}
\usepackage{graphicx}

\begin{document}
\title{Theory of excited state absorptions in phenylene-based 
$\pi$-conjugated polymers}
\author{Alok Shukla}
\affiliation{Physics Department, Indian Institute of Technology, Mumbai 400076,
India}
\author{Haranath Ghosh}
\affiliation{Centre for Advanced Technology, Indore 452013, India}
\author{Sumit Mazumdar}
\affiliation{Department of Physics, University of Arizona, Tucson, AZ 85721}
\date{\today}
\begin{abstract}
%We report the results of large scale multireference singles-doubles 
%configuration 
%interaction (MRSDCI)
%calculations of excited state absorptions in oligomers of 
%poly-(paraphenylene) (PPP) and 
%poly-(paraphenylenevinylene) (PPV) within a rigid-band correlated electron
%model. The goal of these calculations is to 
%develop understanding of the high energy even-parity two-photon states
%that may be observed in ultrafast photoinduced absorption (PA) experiments.
Within a rigid-band correlated electron model for oligomers of 
poly-(paraphenylene) (PPP) and poly-(paraphenylenevinylene) (PPV), we show
that there exist two fundamentally different classes of two-photon
A$_g$ states in these systems to which photoinduced absorption (PA) can occur. 
At relatively lower
energies there occur A$_g$ states which are superpositions of 
one electron - one hole (1e--1h) and two electron -- two hole (2e--2h)
excitations, that are both comprised of the highest
delocalized valence band and the lowest delocalized conduction band states
only. The dominant PA is to one specific member of this class
of states (the mA$_g$). In addition to the above class of A$_g$ states, 
PA can also occur to a higher energy
kA$_g$ state whose 2e--2h component is {\em different} and has significant 
contributions from excitations involving both delocalized and localized
bands. Our calculated scaled energies of the mA$_g$ and the kA$_g$ agree 
reasonably well to the experimentally observed low and high energy PAs in PPV.
The calculated relative intensities of the two PAs are also in qualitative
agreement with experiment. In the case of ladder-type PPP and its oligomers,
we predict from our theoretical work a new intense PA at an energy considerably lower than the region where PA have been observed currently.
Based on earlier work that showed that efficient charge--carrier generation
occurs upon excitation to odd--parity states that involve both delocalized
and localized bands, we speculate
that it is the characteristic electronic nature of the kA$_g$
that leads to charge generation subsequent to excitation to this state, as found
experimentally.
\end{abstract}
\pacs{78.47.+p, 78.20.Bh, 42.65.Re, 78.40.Me, 78.66.Qn}
\maketitle
\section{Introduction}
\label{intro}
The applications of $\pi$-conjugated polymers in optical emission devices such
as organic light emitting
diodes \cite{LED} and laser active media 
\cite{laser1,laser2,laser3} have
led to intensive investigations of photoluminescent materials like PPP and PPV,
their derivatives, and structurally related ladder type materials.
Nonlinear spectroscopy, including
both ultrafast photoinduced absorption (PA) measurements 
\cite{Yan1,Yan2,Leng,Frolov1,Frolov2,Klimov,Kraabel,Silva,Gadermaier1,Gadermaier2} 
and third order nonlinear
optical measurements like electroabsorption (EA) \cite{Leng,Liess,Martin},
third harmonic generation (THG) \cite{THG} and two-photon absorption (TPA)
\cite{Frolov2,Baker,Lemmer} have been carried out extensively to probe the
even parity excited states that are dark under one--photon excitation but
are two--photon allowed. 
In particular, interest in PA experiments stems
from the observation that PA at high energies
is followed by charge separation, with the creation of
charged polarons on neighboring chains. Derivatives of PPV, for example,
have been investigated by Frolov {\em et al.} \cite{Frolov1,Frolov2}, 
who found two distinct PA bands
in these systems, ``low energy'' PA1 and ``high energy'' PA2, occurring 
at $\sim$ 0.8 eV and $\sim$ 1.3 -- 1.4 eV, respectively. Frolov {\em et al.}
ascribe PA1 to the excited state absorption from the optical 1B$_u$ state
to the so-called mA$_g$ state (where m is a chain length dependent
unknown quantum number), 
whose nature has been discussed
extensively by theorists in the context of
nonlinear spectroscopy of both luminescent polymers like PPV as well as
nonluminescent linear chain polyacetylenes and 
polydiacetylenes
\cite{Dixit,Chandross,McWilliams,Soos1,Abe,Beljonne,Yaron,Race}.
PA2 has been ascribed by Frolov {\em et al.}
to the excited state absorption to a higher energy kA$_g$ state (where k is 
again an unknown quantum number),
whose counterpart does not exist in the linear chain polymers, according
to these authors. Interestingly, the relaxation
dynamics of PA1 and PA2 are very different: while the mA$_g$
decays back to the optical 1B$_u$ exciton by internal conversion, the kA$_g$ 
undergoes a different relaxation pathway that leads to dissociation
into long-lived polaron pairs that are probably interchain.
Based on this, the authors have speculated
that the electronic character of the kA$_g$ is different from the mA$_g$. 

Similar behavior have been observed 
by several different groups 
\cite{Klimov,Kraabel,Silva,Gadermaier1,Gadermaier2}, who have studied both
PPV derivatives and structurally related materials like polyfluorene,
ladder-type PPP and oligomers of the latter. 
%%SM1/16 -- deleting following lines
%In all cases 
%strong low energy PA1 and relatively
%weaker high energy PA2 are observed. Irresepective of whether PA2 is a 
%signature of a 
%kA$_g$ state alone \cite{Frolov1,Frolov2}, or whether there occur also 
%contributions
%to PA2 from the polaron absorptions themselves
%\cite{Kraabel,Gadermaier1,Gadermaier2}, the same mechanism of polaron 
%generation
%upon excited state absorption is suggested by these experiments.
%%SM1/16 -- adding new.
There exist apparent subtleties in comparing some of these experimental results
with those of Frolov {\em et al.} \cite{Frolov1,Frolov2},
as the PA measurements in some of these cases were carried out only in the
high energy region ($>$ 1.4 eV) \cite{Gadermaier1,Gadermaier2}, and 
therefore the induced absorption
termed PA1 by these latter authors actually corresponds to PA2 of
Frolov {\em et al.} \cite{Frolov1,Frolov2} (see section V for details). Apart from this difference
in nomenclature, the fundamental observation in all cases appear to be the same,
viz., A$_g$ states beyond some threshold energy mediate interchain charge--transfer.
As an aside, we remark that it has
also been claimed from ultrafast PA measurements in the wavelength region
of infrared active vibrational modes that charge carriers are generated 
directly at the optical threshold, and that the quantum efficiency of charge
generation is wavelength independent \cite{Moses,Miranda}. 
This last conclusion has,
however, been challenged by Silva {\em et al.} \cite{Silva}, 
whose demonstrations that 
ultrafast photogeneration of charge carriers in a polyfluorene derivative
is a consequence of sequential absorption to an A$_g$ state that is
at nearly twice the energy of the optical exciton (note that this is different
from the kA$_g$), and that there occur
negligible yield of polarons under
continuous wave conditions (where excitation to high energy A$_g$ states does
not occur) argue against the direct photogeneration scenario.
The occurrence of EA \cite{Liess,Osterbacka} and TPA \cite{Frolov2} at
the same energy where PA2 occurs are yet other demonstrations of the existence
of a kA$_g$ state
that is dipole-coupled to the 1B$_u$.
Recent photo-current excitation cross-correlation experiments 
\cite{Zenz,Muller} also support the sequential absorption picture. We shall 
therefore make no further comments on the experiments by Moses {\em et al.}
\cite{Moses,Miranda}, which
in any case is of secondary interest of this work; our primary interest is to
understand the different electronic natures of the mA$_g$ and the kA$_g$
at a qualitative level.

Initial progress in theoretical understanding of possibly different classes of
even parity states in
phenylene-based conjugated polymers was made by Chakrabarti and Mazumdar
\cite{Chakrabarti}, who examined the excited states of biphenyl
and triphenyl within the Pariser--Parr--Pople model \cite{PPP1,PPP2}. Although
many--body techniques used by these authors were accurate (exact and
quadruple--CI, hereafter QCI), the basis sets used were limited
and included only the degenerate pairs
of highest occupied molecular orbitals (HOMOs) and lowest unoccupied
molecular orbitals (LUMOs) of benzene. Each pair of degenerate MOs consists of 
a delocalized
MO and a localized MO, with the $\pi$ electron densities vanishing on the
para carbon atoms in the latter \cite{Rice,Cornil,Chandross1,Shimoi}. 
Chakrabarti and Mazumdar showed that 
the two-photon states in these molecules were not only superpositions of
1e--1h and 2e--2h
excitations involving the
delocalized bonding ($d$) and antibonding ($d^*$) MOs,
there occurred also strong admixing with 2e--2h
excitations involving localized bonding ($l$) and antibonding 
($l^*$) MOs. The authors therefore suggested that 
%%SM1/16 - modifying, to make it shorter and less convoluted
%the existence
%of the $l$ and $l^*$ MOs made the higher A$_g$ excited states in phenylene-based
%conjugated polymers qualitatively different from the relatively lower energy
%two-photon states. 
there occurred multiple kinds of A$_g$ states in phenylene--based conjugated polymers.
According to these authors, A$_g$ states upto some
threshold could qualitatively be understood within the space of
$d$ and $d^*$ MOs only. The mA$_g$ belongs to this class of A$_g$ states.
Beyond the threshold, however, there
occurs a crossover to A$_g$ states which have strong contributions from
2e--2h excitations involving $l$ and $l^*$ MOs. We shall hereafter write
such states as $(d \to l^*)^2$ and $(l \to l^*)^2$ double excitations
(as opposed to the $(d \to d^*)^2$ double excitations with lower energies).
Note that in all cases we do not make finer distinctions between configurations
in which the hole or electron pairs occupy the same or different MOs (thus
$(d \to l^*)^2$ includes the 2e--2h excitations $(d \to l^*;l \to d^*)$).
According to these authors, the
kA$_g$ belonged to this second class of A$_g$ states.
No attempt
was made to explain the charge separation and the calculations were limited
strictly to explaining the possible difference between the mA$_g$ and the
kA$_g$. 

The calculations by Chakrabarti and Mazumdar \cite{Chakrabarti} are
suggestive but far from complete. Firstly, 
all calculated A$_g$ states appeared
to have nearly equal admixtures of $(d \to d^*)^2$, $(d \to l^*)^2$, 
$(l \to d^*)^2$ and $(l \to l^*)^2$, which would appear to contradict the
conjecture about different classes of A$_g$ states. It was claimed that the
strong mixing between different classes of 2e--2h excitations was a consequence
of the severe length restrictions in the previous calculations and that with 
increasing conjugation length distinct A$_g$ states dominated by different
kinds of 2e--2h excitations would emerge.
The actual demonstration
of this requires going beyond the earlier small oligomer calculations. 
A second shortcoming of the earlier calculations is that they
completely ignored the outer delocalized MOs of each benzene (as well as
the MOs due to the vinylene segments in case of PPV).
The complete band structure of PPV can be seen, for example, in reference
\onlinecite{Chandross1}, where one finds that in addition to the 
innermost delocalized valence and conduction bands 
(referred to as simply $d$ and $d^*$ in the above, but hereafter as
$d_1$ and $d_1^*$) there occur also outer delocalized bands
($d_2$ and $d_2^*$ ; $d_3$ and $d_3^*$) that lie below (above) the $l$ 
($l^*$) bands (the difference between PPV and PPP is the absence of the
outermost $d_3$ and $d_3^*$ bands in PPP). The neglect of the outer $d$-bands
raises the question whether higher energy PA is not to qualitatively different
A$_g$ states involving the localized levels, 
but to states that are dominated by excitations involving the outer $d$-bands.
Finally, the previous calculations were not useful for even qualitative
comparisons of relative energies and oscillator strengths of theoretical
PA1 and PA2, again a consequence of the very severe length restriction.

In view of the above, we believe
that a thorough and accurate theoretical investigation of the 
two-photon states of
oligomers of PPP and PPV is in order. In the present paper, we present
large scale correlated calculations employing the multireference singles
and doubles
configuration interaction (MRSDCI)
method \cite{Buenker,Tavan} on longer oligomers of PPP and PPV within a  
correlated electron Hamiltonian. Unlike the previous calculations 
\cite{Chakrabarti}, higher energy MOs or bands are not ignored.
We take considerable care that the results we report are accurate, by imposing
strict convergence criteria on our MRSDCI calculations. 
For PPP, calculations are presented
for oligomers containing  three to five benzene rings. We refer
to these as PPP3, PPP4 and PPP5, respectively.  
In the case of
PPV, we consider oligomers that terminate with benzene molecules at both
ends, in order to preserve spatial symmetry, and present 
MRSDCI results 
for oligomers containing three (PPV3) and four (PPV4) benzene
rings. While these oligomers are still relatively short, we believe that it is
far more important to incorporate electron correlation effects to high order
than to go to longer oligomers. 

The results of our calculations can be summarized as follows. Firstly,
our computed PA spectra 
%%SM1/16 -- rearrange
%in all cases
resemble the experimental PA spectra qualitatively: 
%in
%the sense that 
in all cases we find the theoretical PA1 to be much stronger than PA2.
We discuss why the calculated A$_g$ states are expected to
occur at energies where PA1
and PA2 are observed experimentally.
Secondly, our present relatively longer chain calculations do indeed find 
lower and higher energy two--photon states whose 2e--2h components are
significantly different, as would be required to claim that these belong
to distinct classes.
%%SM1/16 -delete 
%As we remark later, 
The different
nature of the higher energy two--photon state that we 
%SM1/16 - change wording: assign 
believe to be the
kA$_g$ may explain the charge separation
from this state. 
While this last statement is a speculation currently, it
is supported by earlier photocurrent studies \cite{Kohler}.
A third result involves our determination that the mA$_g$ is not the lowest
two-photon excited state 
in any of the systems we have studied (i.e., the
quantum number $m >$ 2), even though the 2A$_g$ occurs above
the 1B$_u$ in these systems. This particular 
result is in agreement with other recent theoretical
work based on the limited basis of $d_1$ and $d_1^*$ bands 
\cite{Lavrentiev,Beljonne1}.
As we remark later, this may also have 
experimental significance. Finally, we also find that as observed before
from fittings of the linear absorption in PPV films \cite{Chandross2}, 
the standard
Ohno parameters \cite{Ohno}
may be too large  
for even qualitative
fittings of the PA at high energies, 
and a previously used phenomenological screened
Coulomb parametrization \cite{Chandross2} gives more satisfactory 
results.

In the next section we present our theoretical model  
and discuss the details of the MRSDCI approach as adopted here.
In section III we present a brief review of the theory
of linear absorption in the polyphenylenes, to point out that there occur
distinct classes of one--excitation in the systems of interest, thereby
suggesting the idea of distinct two--excitations. In section IV we present
our theoretical results. This is followed
by our conclusions and discussions of the scope of future work.  

\section{The theoretical model and methodology}
\label{model}

We consider oligomers of PPP and PPV within the Pariser-Parr-Pople model
Hamiltonian,
\begin{eqnarray}
\label{H_PPP}
H = - \sum_{\langle ij \rangle, \sigma} t_{ij}
(c_{i \sigma}^\dagger c_{j\sigma}+ c_{j \sigma}^\dagger c_{i \sigma}) + \nonumber \\
U \sum_i n_{i \uparrow} n_{i \downarrow} +
\sum_{i<j} V_{ij} (n_i -1)(n_j -1)
\end{eqnarray}
\noindent where $\langle ij \rangle$ implies nearest neighbors,
$c_{i \sigma}^\dagger$ creates an electron of spin
$\sigma$ on the $p_z$
orbital of carbon atom $i$, $n_{i \sigma} = c_{i \sigma}^\dagger c_{i \sigma}$
is the number of electrons with spin $\sigma$, and
$n_i = \sum_\sigma n_{i \sigma}$ is the total number of electrons on atom $i$.
The parameters $U$ and $V_{ij}$ are the on--site
and long--range Coulomb interactions, respectively, while $t_{ij}$ is the
nearest neighbor one-electron hopping matrix element that includes
bond alternation and connectivity. The parametrization of the intersite Coulomb
interactions is done in a manner similar to the Ohno
parametrization \cite{Ohno}
\begin{equation}
V_{i,j} = U/\kappa (1+0.6117R_{i,j}^2)^{1/2} \; \mbox{,}
\label{eq-ohno}
\end{equation}
where $\kappa$ is a parameter which has been introduced to account for
the possible screening of the Coulomb interactions in the
system \cite{Chandross2,Castleton}. We have examined both the standard Ohno parameters
($U$ = 11.13 eV, $\kappa$ = 1.0), as well as a
particular combination of $U$ and $\kappa$
($U$ = 8.0 eV, $\kappa$ = 2.0) that was shown previously to be
satisfactory at a semiquantitative level
for explaining the full wavelength dependent ground state
absorption
spectrum of PPV \cite{Chandross2}.
We shall hereafter refer to this second set of parameters as screened Ohno
parameters. As far as the hopping matrix elements are
concerned, we took $t = -2.4$ eV
for the C -- C bond in benzene rings.
The hopping corresponding to the interbenzene single bond in the PPP oligomers
was taken to be $t=-2.23$ eV. For the vinylene linkage of the PPV oligomers,
we chose the hopping elements to be $-2.2$ eV for the single bond, and $-2.6$
eV for the double bond.
We considered PPP and PPV oligomers in their planar configurations, with
the conjugation direction along the $x$ axis. Thus
the symmetry group of PPP oligomers is $D_{2h}$, while that
of PPV oligomers is $C_{2h}$. Since both symmetry groups have inversion
as a symmetry element, the many electron states of these oligomers can be
classified according to this symmetry. 
The two-photon states of both the
PPP and PPV oligomers belong to the A$_g$ irreducible representation (irrep)
of the respective symmetry group, while
the one-photon states belong to the B$_u$ irrep for PPV, and B$_{1u}$
($x$-polarized), and
B$_{2u}$ ($y$-polarized) irreps for the oligo-PPPs. Since, here we are
concerned mainly with the response of these systems to the 
$x$-polarized photons,
therefore, henceforth, for PPP oligomers also we will refer to the B$_{1u}$
states as B$_u$ states.
In all cases the calculated PA corresponds to the excited state absorption
from the optical 1B$_u$ state. 
In all the many-body calculations presented
in this work, full use of the stated point group symmetries was made.

As stated above, the correlated electron calculations were done using the
MRSDCI approach, which is a powerful CI technique \cite{Buenker} that
has been used previously for linear chain polyenes by Tavan and Schulten
\cite{Tavan} as
well as others \cite{Beljonne}, and by us to calculate the excited state
ordering in polyphenyl- and polydiphenylacetylenes \cite{Ghosh}. As discussed
in reference \onlinecite{Ghosh} we use very stringent convergence criterion
for all excited states.
The methodology behind the MRSDCI calculations is as follows.
The calculations are initiated with a restricted Hartree-Fock 
(RHF) computation of the
ground state of the oligomer concerned, followed by
a transformation of the Hamiltonian from the site representation 
(Eq.~\ref{eq-ohno}) to the
HF molecular-orbital representation. 
Subsequently, a singles-doubles CI (SDCI) calculation
is performed, the different excited states in A$_g$ and B$_u$ subspaces are
examined, and the N$_{ref}$
configuration state functions (CSFs) making
significant contributions to their many-particle wave functions are
identified.  The next step
is the MRSDCI calculation for which the reference space consists of the
N$_{ref}$ CSFs identified in the previous step,
and the overall Hamiltonian matrix now includes configurations that are 
singly and doubly
excited with respect to these reference CSFs (thereby including the
dominant triply and quadruply excited configurations).
The new ground and excited states are now re-examined to identify new CSFs 
contributing
significantly
to them so as to augment the reference space for the next set of MRSDCI
calculations. This procedure is repeated until
satisfactory convergences in the excitation energies of the relavant
states are achieved.
By the time convergence is achieved
typically all CSFs with coefficients of magnitude
$0.1$ or more in the corresponding many-particle wave functions have been
included in the MRSDCI reference space. Naturally, this leads to very large
CI matrices. To give some idea of the highly correlated nature of the
wavefunctions of the excited states we have examined and the level of accuracy 
in our calculations, in Table \ref{tab-nref} we have listed the number of 
reference
functions that were used for each symmetry subspace of PPP3, PPP4 and PPP5,
respectively, as well as the overall sizes of the Hamiltonian matrix in each
case (note that for PPP3 the method used was QCI rather than MRSDCI). The
number of MRSDCI reference functions are larger in the A$_g$ subspaces than in 
the B$_u$ 
subspace because while only the 1B$_u$ was optimized in the B$_u$ subspace,
many different A$_g$ states (all those with significant transition dipole
couplings with the 1B$_u$) had to be simultaneously optimized in the
A$_g$ subspace. The $N_{ref}$ in Table \ref{tab-nref} should be compared to the few (usually
2 or 3) reference states that are retained in calculations of the lowest A$_g$
states \cite{Tavan}. To the best of our knowledge, the present calculations are
the most accurate correlated electron calculations that incorporate the full 
basis set for the polyphenylenes.
\section{Theory of ground state absorption and its implication}
\label{review}

Before we present our calculations of PA, 
it is useful to recall the results
of calculations of the ground state absorption \cite{Rice,Cornil,Shimoi,Chandross1,Chandross2}. 
This is because multiple classes of final states are relevant also in
ground state absorption, and as we indicate below analysis of the ground state
absorption strongly suggests that there should occur multiple classes of
two-photon states.
Within band theory ($U$ = 0) ground state absorption consists of 1e--1h
excitations that are low energy $d_1 \to d_1^*$, high energy $l \to l^*$,
and intermediate energy $d_1 \to l^*$ and $l \to d_1^*$, with the intermediate
energy absorption band occurring exactly half way in between the low and high
energy absorption band. Here we have ignored absorptions involving $d_2$, 
$d_2^*$ etc. bands, as excitations involving these bands lie outside the range
of experimental wavelengths.  
The $d_1 \to d_1^*$ and $l \to l^*$ bands are polarized
along the $x$-direction in PPP and predominantly along the $x$-direction in
PPV, while the $d_1 \to l^*$ and $l \to d_1^*$ bands are polarized along
the $y$-direction and predominantly along the y-direction, respectively. 
Experimentally in PPV 
there occur absorptions at $\sim$ 2.4 eV,
3.7 eV, 4.7 eV and $\sim$ 6.0 eV, respectively. These have been explained
within a correlated electron picture: 
absorptions at 2.4 eV
and 6.0 eV are due to $d_1 \to d_1^*$ and $l \to l^*$ exciton states, 
respectively \cite{Rice,Cornil,Chandross2,Shimoi} 
the absorption at 3.7 eV is to a higher energy $d_1 \to d_1^*$ 
exciton \cite{Chandross2}; 
and finally, the absorption at 4.7 eV is
to the ``plus'' linear combination of the excitations 
$d_1 \to l^*$ + $l \to d_1^*$ \cite{Rice,Cornil,Chandross2,Shimoi}. 
The corresponding ``minus'' linear combination,
$d_1 \to l^*$ -- $l \to d_1^*$, occurs also at about 3.7 eV
\cite{Rice,Cornil,Chandross2,Shimoi}, but is forbidden
in linear absorption.
All of these assignments have been confirmed
by polarization studies of absorptions in stretch--oriented samples
\cite{Chandross2,Comoretto,Miller}. 

The relevance of these known results to the present case are as follows.
Energetically, the kA$_g$ in PPV is at $\sim$ 3.6 -- 3.8 eV 
\cite{Frolov1,Frolov2}. Since ground state absorption occurs to a high energy
$d_1 \to d_1^*$ excitation of B$_u$ symmetry in this energy region,
in principle, the kA$_g$
can simply be a similar $d_1 \to d_1^*$ excitation of 
A$_g$ symmetry. If this were true, the mA$_g$ and
the kA$_g$ would be qualitatively similar,
and charge carrier 
creation from the kA$_g$ (but not from the mA$_g$) can only be a 
consequence of possibly greater e--h separation in the kA$_g$.
An alternate possibility is that the kA$_g$
is dominated by configurations that are fundamentally different. 
Recall that 
the occurrence of the 2A$_g$ below the 1B$_u$ in polyacetylenes and
polydiacetylenes is a general many--body phenomenon:
A$_g$ eigenstates having strong 2e--2h contributions from specific
MOs can be close in energy, or can even occur below B$_u$
eigenstates that are dominated by 1e--1h 
excitations involving the same MOs. Since the lowest energy 1e--1h excitations
involving the $d_1 \to l^*$ and $l \to d_1^*$ excitations (the minus linear
combinations mentioned above) occur at $\sim$ 3.7 eV,
it is to be anticipated that 
many-body A$_g$ eigenstates that are dominated by $(d_1 \to l^*)^2$ and 
$(l \to d_1^*)^2$ occur also within the same energy range. In principle then,
the kA$_g$ can also be dominated by 2e--2h components involving the $l$ and
$l^*$ bands, and be qualitatively different from the mA$_g$. As shown in the
following sections, we do indeed find evidence for such A$_g$ states with
significant dipole coupling to the 1B$_u$.

\section{Results of correlated electron calculations}

\subsection{Choice of parameters}

As we show below electron correlation effects on different kinds of CSFs are
different. This necessitates proper
choice of the Coulomb parameters in Eq.~\ref{H_PPP}.
Here we show that the bare Ohno parameters are not suitable for high energy
states of PPP 
and PPV. In Fig. 1 we show the calculated excited state 
absorption from the 1B$_u$ state for 
PPP3, for the case of Ohno parameters, obtained with the QCI approach.  
The spectral features I and II are due to the 2A$_g$ and the
4A$_g$ states of the oligomer, respectively,
while the strong feature III has contributions from
both the 7A$_g$ and the 8A$_g$. 
The close proximity of 
spectral feature I
to the 1B$_u$ in the theoretical spectrum makes this outside
the wavelength range within which the experimental PA features are observed.
In principle, it might be possible to observe this
2A$_g$ state in TPA. In reality, Fig.~1 indicates that the
strength of the TPA to the 2A$_g$ would be rather weak. This is because the
calculated PA in Fig.~1 corresponds to linear absorption from the
1B$_u$ and hence consists of only positive terms; in contrast, the 
third order susceptibility corresponding to TPA contains both positive and 
negative terms \cite{FGuo}, and the relative strength of the 2A$_g$ peak in
Fig.~1 therefore corresponds to an upper limit for TPA. 
Thus only calculated
PA features II and III
in Fig. ~1 should be compared to experiments. The relative
oscillator strengths of II and III are exactly opposite to the relative
oscillator strengths of PA1 and PA2, which makes the calculated PA 
inconsistent with experiments \cite{Frolov1,Frolov2}.

In order to probe this further, we have examined the final states
 of all PA features 
in considerable detail. Each of these wavefunctions is highly correlated, and 
is a superposition of numerous configurations. In order to obtain broad
classifications of the different A$_g$ states 
we expand the correlated wavefunction as, 

\begin{eqnarray}
\label{wavefunction}
|nA_g \rangle = \sum_i a_i |d_1 \to d_1^* \rangle_i + \sum_j b_j |(d_1 \to d_1^*)^2 \rangle_j \nonumber \\
+ \sum_k c_k |(d_1 \to l^*)^2 \rangle_k + ....
\end{eqnarray}

In the above nA$_g$ is an arbitrary A$_g$ state and each term on the right hand
side contains all CSFs of a given class (for example,
$|d_1 \to d_1^* \rangle_i$
is the $i$th configuration of the type $d_1 \to d_1$ whose coefficient in 
nA$_g$ is $a_i$). 
The right hand
side of Eq.~\ref{wavefunction} is obviously not complete and the .... implies
the existence of many other types of excitations $d_1 \to d_2^*$,
$(d_1 \to d_2^*)^2$, etc. Once terms have been collected in the above manner,
it is possible to quantify the overall contribution of excitations of a given
kind (for example, $\sum_i |a_i|^2$ is the total contribution by excitations
of the type $d_1 \to d_1^*$). 
In Table II we have
given the contributions of each kind of excitation
that describe the 2A$_g$, 4A$_g$, 7A$_g$ and 8A$_g$ wavefunctions. 
The overall sum of
the contributions corresponding to each A$_g$ state
does not add up to 1, since we have retained only the 
important excitations with coefficients at least 0.1 in our 
expansions of the wavefunctions. 
As seen from Table II, the 2A$_g$ wavefunction is very 
similar to the 2A$_g$ of linear chain polyenes, in that it is 
predominantly a superposition
of 1e--1h $d_1 \to d_1^*$ and 2e--2h $(d_1 \to d_1^*)^2$ excitations. This
supports the earlier application of the effective linear chain model for the
description of this state in PPV and PPP \cite{Soos2}.
The two states 7A$_g$ and 8A$_g$ also have strong contributions from
$(d_1 \to d_1^*)^2$, and weak contributions from $(d_1 \to l^*)^2$ or
$(l \to d_1^*)^2$. These states are therefore physically related to
the mA$_g$ that has been discussed before in the context of nonlinear
spectroscopy. 
In contrast to the above states, 
the 4A$_g$ has much stronger contribution
from $(d_1 \to l^*)^2$ and $(l \to d_1^*)^2$.
The occurrence of a $(d_1 \to l^*)^2$ and $(l \to d_1^*)^2$ type state 
{\it below} the
predominantly $(d_1 \to d_1^*)^2$ states suggests that (a) electron correlation
effects are stronger on CSFs involving the 
localized MOs than on CSFs involving only delocalized MOs, and 
(b) the bare Ohno Coulomb
parameters are too strong to describe PPP or PPV films, since it is the large
magnitude of the Ohno correlation parameters that
causes the effective crossing of eigenfunctions of different types, which in
turn leads to the reversal of the intensities of the
absorption bands as a function of energy.

Although in the above we have shown the theoretical PA spectrum only
for PPP3, we emphasize that identical behavior is seen with the Ohno parameters
for PPP4, PPP5, PPV3 and PPV4. In all cases the intensity profile of the
excited state absorption is opposite to that observed experimentally, and
wavefunction analysis indicates that this is due to the ordering of the states
as in the above. 

That the Ohno parameters are too large for high energy states of
PPV, PPP etc. was suggested earlier from calculations of ground state
absorption of PPV \cite{Chandross3}.
With the 
Ohno parameters, the calculated $l \to l^*$ absorption (polarized along
the $x$-direction) occurs below the
calculated $(d_1 \to l^* + l \to d_1^*)$ absorption ($y$-polarized)
\cite{Chandross3},
in contradiction to polarized absorption experiments 
\cite{Chandross2,Comoretto,Miller}. 
As pointed out in reference \onlinecite{Chandross3}, this last 
theoretical result 
was obtained even for trans-stilbene in one of the earliest
computational works \cite{Beveridge}. Very recently, Castleton and Barford
\cite{Castleton}
have done careful analysis of 
a very large number of high energy excited states in benzene, biphenyl and
trans--stilbene using full--CI and have reached the same conclusion. 
Within Eq.~\ref{eq-ohno},
the authors suggest $U$ = 7.2 eV and $\kappa$ = 1.36 for the hydrocarbon
matrix condensed phases of the above molecules, along with finer modifications
of the hopping integrals. These values of $U$ and $\kappa$ are close to the
$U$ = 8 eV and $\kappa$ = 2.0 that were suggested in reference 
\cite{Chandross3}, and that we use in our calculations in the next section.
Given that our goal is semiquantitative only 
(recall that our calculations are for
relatively short oligomers while the available experimental results 
include those for
much longer chains) finetuning of the parameters as done by
Castleton and Barford would be premature. The necessity to incorporate
screening of the bare Ohno Coulomb parameters in order to fit the high energy
excited states in condensed phases of conjugated polymers have also been
discussed by Moore and Yaron \cite{Moore}, from a different perspective. 

\subsection{PA spectra with screened Coulomb parameters}

In Fig.~2 we have shown the calculated excited state
absorptions for (a) PPP4,
(b) PPP5, (c) PPV3 and (d) PPV4 for $U$ = 8.0 eV and $\kappa$ = 2. 
In all cases our abcissa is the energy
scaled with respect to E(1B$_u$). As pointed out before \cite{DGuo},
convergence in the scaled energies of high energy states 
with increasing chain length is not expected in this region of relatively 
short chain lengths. This is because the lowest energy excitations 
(for example, the 1B$_u$ and the 2A$_g$) converge with increasing chain length
much faster than the higher energy states. Taken together with the discrete
nature of the excited states, this can indicate an {\it apparent} increase
in the scaled energy of the high energy excited state with increasing
chain length, even as the actual energy is decreasing. This is exactly what
happens between PPP4 and PPP5, and between PPV3 and PPV4 in 
Fig.~2.
Quantitative comparisons of the scaled
energies of the theoretical excited state absorption bands and those of the
experimental PA bands thus cannot be expected, particularly
in view of the fact that the relative location of the A$_g$ state
dominated by $(d_1 \to l^*)^2$ excitations is very sensitive to
relatively minor changes in the Coulomb correlation parameters 
(see previous subsection and also below). The energies and the relative 
intensities of the theoretical induced absorptions are therefore for
semiquantitative comparisons only. 
From Figs. 2(c) and 2(d), it can
be concluded that PA1 in PPV derivatives is due to the calculated
spectral feature II which peaks slightly below 0.4 $\times$ E(1B$_u$), which
would correspond to about 0.9 eV in long chain PPV derivatives (with
E(1B$_u$) $\sim$ 2.2 eV).
This is quite close
to experimental PA1 energy in PPV derivatives \cite{Frolov1,Frolov2}. 
We remark on the PA energies in PPP derivatives
in the next section, where more detailed comparisons to experiments are made.
In all cases, the theoretical PA feature I
corresponds to the 2A$_g$ (see below), and the close proximity of this state
to the 1B$_u$ again suggests that this state is outside the wavelength region
within which experimental PA has been observed. 

\begin{figure}
\includegraphics[height=7cm, width=8cm]{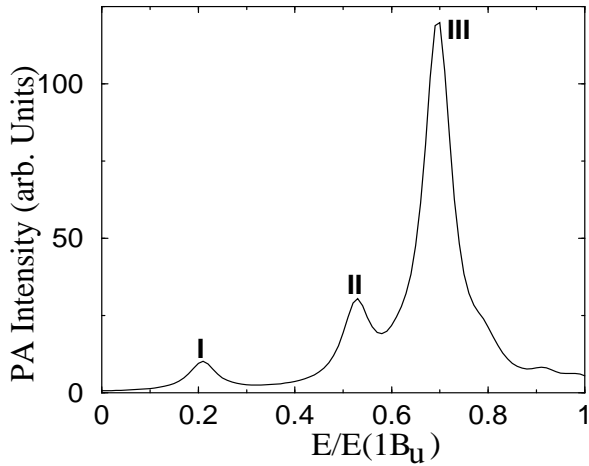} 
\caption{Calculated PA spectrum of PPP3, with the standard
Ohno Coulomb parameters. A linewidth of 0.15 eV was assumed.}
\label{PPP3-ohno}
\end{figure}
For detailed understanding,
in Table III we have described the wavefunctions of each of the 
final A$_g$ states corresponding to each band in the calculated PA spectra. 
The items
in Table III are similar to those in Table II, i.e., each entry corresponds to
the overall contribution of each kind of an excitation. Remarkably, the
wavefunction descriptions of the different spectral features are very
similar for all four systems shown in Fig. 2. Broadly 
speaking, there occur three distinct
classes of A$_g$ states in all four systems. These are discussed below.

\begin{table}
\caption{The number of
reference
configurations ($N_{ref}$) and the total number of configurations ($N_{total}$)
involved in the MRSDCI (or QCI, where indicated) calculations, for different
symmetry subspaces of the  various oligomers.}
 
\begin{tabular}{lrlrr} \hline \hline
Oligomer & \multicolumn{2}{c}{$A_g$} &  \multicolumn{2}{c}{$B_u$} \\
     &  $N_{ref}$ & $N_{total}$ &  $N_{ref}$ & $N_{total}$ \\ \hline \hline
PPP3 & 1$^a$ & 193678 & 1$^a$ & 335545 \\
PPP4 & 55    & 284988 & 15 & 76795 \\
PPP5 & 48    & 663619 & 7 & 87146 \\
PPV3 & 37    & 215898  & 12 & 220905 \\
PPV4 & 39   & 981355  & 3  & 225970  \\ 
\hline \hline
\end{tabular}
 
{\noindent $^a$ QCI method}
\label{tab-nref}
\end{table}                                                            
The first class of states is represented by the 2A$_g$, which in all cases
is predominantly a superposition of 1e--1h $d_1 \to d_1^*$ and 2e--2h
$(d_1 \to d_1^*)^2$, as with the Ohno parameters. 
However, compared to the bare Ohno 
parameters, the relative weight of the 1e--1h excitations here is larger. This
is a conseqeuence of the smaller $U$, and is to be expected. Furthermore,
the relative weight of the 1e--1h excitations is also larger in the PPP
oligomers than in the PPV oligomers. This is also to be expected, based on the
larger one--electron gap in PPP. The 2A$_g$ can certainly be descibed within
an effective linear chain model with large dimerization that retains only the
$d_1$ and $d_1^*$ bands, as suggested before \cite{Soos2}.

\begin{table}
\caption{Relative weights of the dominant contributions to the excited states 
of PPP3, computed with the standard Ohno parameters (see text).} 

\begin{tabular}{lccccc} \hline \hline
PA Feature & State & $d_1 \rightarrow d_1^{*}$ & 
  $d_1 \rightarrow d_2^{*}$ & $(d_1 \rightarrow d_1^{*})^2$ & 
$(d_1 \rightarrow l^{*})^2$  \\ \hline \hline
 I   & $2A_g$   & 0.3585 & 0.0601 & 0.2670  &  --- \\
 II  & $4A_g$   & 0.0452  & 0.1521 & 0.1266  & 0.2282 \\ 
 III & $7A_g$   & 0.1116  & 0.1280  & 0.1152 &  0.0660 \\
     & $8A_g$   & 0.0939 & 0.0772 & 0.3155  & 0.0985  \\
% IV  & $12A_g$   & ---    &  &  &  \\
%  V  & $15A_g$  &  ---   &  &  &  \\ 
\hline \hline 
\end{tabular}
\label{tab-ppp3}
\end{table}
The second class of states are represented in all cases by the different
A$_g$ states that form the final states in the
second and third bands in the calculated PA spectra. These have
strong contributions from 2e--2h $(d_1 \to d_1^*)^2$, and weak but nonzero
contributions from $(d_1 \to l^*)^2$
and $(l \to d_1^*)^2$. 
In addition, there occur also 1e--1h contributions of the
type $d_1 \to d_2^*$, $d_2 \to d_1^*$, etc., and while there are subtle 
differences in the relative contributions by different kinds of single
excitations involving low and high energy delocalized bands, the overall
natures of the excitations that are the final states in absorption bands II
and III are similar. The qualitative natures of these eigenstates are very
similar to that of the mA$_g$ discussed in the context of nonlinear 
spectroscopy of polyacetylenes and polydiacetylenes. The wavefunction
descriptions make it clear that very qualitatively these wavefunctions
can also be described within the effective linear chain model,
but with less precision than the 2A$_g$. Our calculated
relative intensities of the 2A$_g$ and the mA$_g$ are in qualitative
agreement with 
other recent calculations for polyphenylenes that used a basis space of
only the $d_1$ and $d_1^*$ bands \cite{Lavrentiev,Beljonne1}.  

\begin{figure}
\includegraphics[height=7cm, width=8cm]{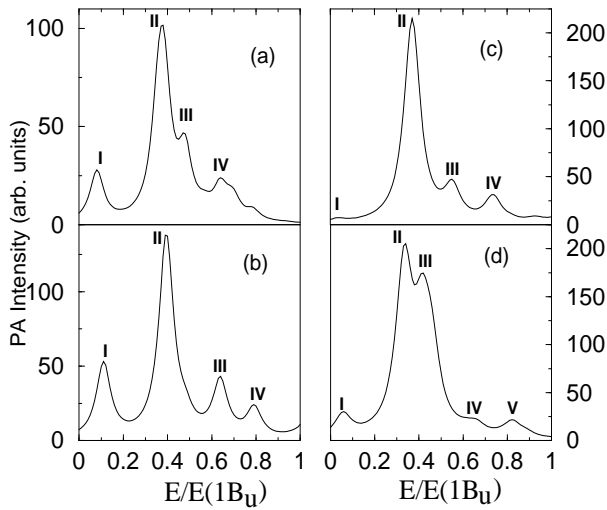} 
\caption{Calculated PA spectra of (a) PPP4, (b) PPP5, (c) PPV3, and
(d) PPV4, with screened Coulomb parameters. The scales for the
intensity are different in different
cases. A line
width of 0.15 eV was assumed in all cases.}
\label{PA-spectra}
\end{figure}
Above band III in the calculated spectra we always have a state that is 
qualitatively different from lower energy A$_g$ states. 
%%SM1/16 -- Alok please check next sentence, -- this is new. 
%% Is my calling the appropriate feture in PPV4 as
%% V correct? There was a funny line (item (iv)) in your email of Jan 14 that I 
%% did not understand.
This corresponds to the
spectral feature IV in PPP4, PPP5 and PPV3, and the spectral feature V in PPV4.
The relative weights
of the two classes of 2e--2h excitations, $(d_1 \to d_1^*)^2$ and
$(d_1 \to l^*)^2$, are reversed in this high energy state, with the
former now making weak contribution and the latter a strong contribution.
This qualitative difference between the A$_g$ states at different energies
is in agreement with the earlier conjecture by Chakrabarti and Mazumdar
\cite{Chakrabarti}. The clear demarkation between
A$_g$ states dominated by either $(d_1 \to d_1^*)^2$ or $(d_1 \to l^*)^2$,
rather than nearly equal admixing, is a new result.
This is an effect of increased oligomer length, as discussed above. Furthermore,
we also notice that the highest energy two--photon states in Fig. 2 have the
strongest 1e--1h contribution from excitations of the type $d_1 \to d_2^*$ 
(in the case of PPP oligomers) and $d_1 \to d_3^*$ (in the case of PPV oligomers).
This is yet another difference between the mA$_g$ and the kA$_g$.
What is also significant is the weak role of $(l \to l^*)^2$ 2e--2h excitations
in the A$_g$ states included in Table III, in contrast
to their relatively strong role in the high energy A$_g$ states of
biphenyl and triphenyl within the previous limited basis calculations
\cite{Chakrabarti}. This is also an effect of increased chain length,
with the $(l \to l^*)^2$ 2e--2h excitations now perhaps dominating even
higher energy distinct two-photon states. 
The weaker oscillator strength of absorption
band IV, relative to the strong oscillator strength of absorption band II
is easily understood from Table III: 2e--2h excitations of the type
$(d_1 \to l^*)^2$  have small dipole coupling with 
the 1B$_u$, which is predominantly $d_1 \to d_1^*$. Thus the dipole coupling
of the A$_g$ states responsible for absorption band IV originates mostly
from the small 1e--1h contribution to this state. These highest energy
A$_g$ states are neither expected nor found in the limited basis calculations
by Lavrentiev {\em et al.} \cite{Lavrentiev} and Beljonne \cite{Beljonne1}.

Taken together, the results of Table III indicate the occurrence of 
different classes
of A$_g$ states, which are distinguished by the dominance of different types
of 2e--2h excitations. With hindsight, this is perhaps not entirely
surprising, as discussed in section~\ref{review}. 

\section{Conclusions and Discussions}

%%SM116 -- modifying this paragraph, pl. read carefully, particularly last sentence.
Even as the concept of different types of two--photon states appear to be
correct, it might seem that
straightforward assignments of experimental PA1 and PA2 from the 
calculated PA
spectra of Fig.~2 alone are not possible. We first discuss applications
of our theory to PPV derivatives, and then to ladder PPP, etc. The strong spectral
feature II in the theoretical spectra (the theoretical mA$_g$) is certainly
a component of the experimental PA1 in PPV derivatives. Beyond this there are  
two possibilities, viz., (i) the spectral feature III corresponds to PA2, and
the spectral feature IV (V in PPV4) 
is too high in energy to be observed experimentally;
or (ii) the spectral feature III is also a part of PA1 (especially in long
chains) and it is the high energy $(d_1 \to l^*)^2$ excitation
that corresponds to PA2. We ascribe the $(d_1 \to l^*)^2$ excitation
to PA2 based on the following reasons. First, A$_g$ states that
give rise to spectral feature III occur also in linear chain polyenes
\cite{Chandross}. Their contribution to EA and TPA are vanishingly weak in long
chains. In
contrast, the contributions of the experimental kA$_g$ to EA and TPA in
PPV derivatives are clearly visible \cite{Frolov2,Liess,Osterbacka}. On the
other hand, eigenstates ithat are predominantly  $(d_1 \to l^*)^2$
are clearly absent in
polyacetylenes and polydiacetylenes, which possess only delocalized valence
and conduction bands, and hence our 
assignment would naturally explain the absence of the kA$_g$ state in
these systems \cite{Liess,Osterbacka}. Second, as we have already remarked,
spectral feature III (but not IV) has also been found in the calculations by 
Beljonne \cite{Beljonne1}, who, however, determined that in the long chain 
limit the energies of features II and III converge. This once again supports
our assignment of the $(d_1 \to l^*)^2$ excitation to PA2 (note also that the
1e--1h contributions of the highest energy two--photon states are different,
further justifying the notion that these states belong to a different class from
the lower energy 2A$_g$ and the mA$_g$). 
We have already pointed out that the calculated
PA1 energy is reasonably close to the experimental PA1 energy in PPV 
derivatives.
The calculated scaled energy of PA2 from Figs. 2(c) and (d),
at 0.7 -- 0.8 $\times$ E(1B$_u$) (again with experimental E(1B$_u$) = 2.2 eV in substituted
PPV's) is at 1.5 -- 1.76 eV, which is a reasonably good fit to the experimental PA2 energy
of 1.3 -- 1.4 eV, given the very short lengths of our oligomers, and the 
increased difficulty of fitting of very high energy states. 

As mentioned in section I, we believe that the recent experiments on a methyl--substituted
ladder type PPP ($m$-LPPP) \cite{Gadermaier1}
and on a ladder type oligophenyl \cite{Gadermaier2} have probed an energy
region that is considerably above the region where the mA$_g$ occurs. The PA features
that have been called PA1 in these works occur at 1.5 eV in the polymer and at 1.8 eV in 
the oligomer, and are therefore too high in energy to be the same as PA1 in PPV derivatives.
We therefore believe that the observed lowest energy PA features actually correspond to PA2 
of Frolov {\em et al.}, and this is why charge separation occurs upon excitation to this energy.
This assignment is supported by the observation of EA \cite{Harrison1}
as well as TPA \cite{Harrison2} to a different lower
energy two--photon state that occurs $\sim$ 0.7 eV above the 1B$_u$ exciton in the
polymer. We therefore make the testable prediction that there should occur in 
these systems a 
lower energy PA at $\sim$ 0.7 eV that is considerably 
stronger than the PAs in the 1.5 -- 1.8 eV range.  
The scaled PA2 energy from our PPP oligomer calculations,
0.6 -- 0.8 $\times$ E(1B$_u$), with experimental E(1B$_u$) = 2.7 eV in
$m$-LPPP \cite{Harrison1},
corresponds to 1.6 -- 2.1 eV, once again reasonably close to the experimental 
value.

The above then leads to the question why dissociation of the kA$_g$ to 
polaron pairs is so efficient. We speculate that this is related to the
specific structure of the kA$_g$. 
In photoconductivity measurement on a PPV
derivative it has been
found that a large jump in the photoconductivity occurs at 4.7 eV, exactly 
where the $d_1 \to l^* + l \to d_1^*$ one-photon exciton is located 
\cite{Kohler}. 
The jump in the photoconductivity is due to a sudden increase in
the interchain charge carrier generation subsequent to the excitation to this
particular excited state. We speculate that there occurs a similar
enhanced charge carrier generation subsequent to the sequential excitation
to an A$_g$ state that has strong contributions from $(d \to l^*)^2$ 
(whose energy
is however lower and close to the minus combination of the one--excitations).
Indeed, similar mechanism has also been suggested by Zenz {\em et al.} \cite{Zenz}.
In the original work by K\"ohler {\em et al.}, theoretical calculations suggested
that the enhanced dissociation of the $d_1 \to l^* + l \to d_1^*$ state
was a consequence of the highly delocalized electron--hole character of
this state \cite{Kohler}. The latter necessarily
implies that the charge carriers are
similarly weakly bound even in the $(d_1 \to l^*)^2$ double excitation.
Additional contribution to the 
enhanced tendency
to charge separation in the case of PPVs from excitations 
involving $l$ and $l^*$ bands may also
come from phenyl ring rotations that occur in these excited states.
Given that the charge densities on the para carbons are exactly zero for the
$l$ and $l^*$ MOs, bond orders involving the para carbon atoms
in the kA$_g$ will be particularly small.
This might lead to greater phenyl ring 
rotation in the kA$_g$ than in the 1B$_u$ excitation. 
This feature of the
kA$_g$ can lead to a lifetime that is longer than the mA$_g$, and it is conceivable
that the relatively long lifetime and the weak binding between the electrons and holes
together contribute to the charge separation.
In a recent work, one of us (S.M.)
and colleagues have calculated the relative yields of singlet and triplet
excitons starting from oppositely charged polarons, in the presence of
interchain hopping of electrons and holes \cite{Wohl,Tandon}. 
The computational technique used to calculate charge recombination in
these works
can be applied also to the opposite process of photoinduced
charge transfer. Photoinduced charge transfer from different
classes of A$_g$ states is of interest in the present context and is currently
being investigated.

Our observation that the 2A$_g$ is not the mA$_g$ may also have experimental
significance.
An early measurement
detected strong two-photon 
fluoroscence in PPV from a state that is only 0.5 eV above the 1B$_u$ 
\cite{Baker}, and evidence for a second two--photon state slightly higher in energy.
 Later experiments involving nonlinear absorption have 
invariably found
the mA$_g$ to be at least 0.8 eV above the 1B$_u$. It is conceivable that the low energy
two--photon state found by Baker {\em et al.} is the 2A$_g$.
A similar suggestion has also been 
made by Lavrentiev {\em et al.} \cite{Lavrentiev}.

\section{Acknowledgements}

Work at Arizona was partially supported by NSF DMR--0101659, 
NSF ECS--0108696, and the ONR. We 
acknowledge many useful 
discussions with G. Lanzani and Z.V. Vardeny.

\pagebreak

\begin{table}
\caption{Relative weights of the dominant contributions to the excited states of
different oligomers of PPP and PPV, computed with the screened Coulomb
parameters (see text). Note that the $d_3$ band does not occur in PPP}
\begin{tabular}{lccccccc} \hline \hline
Oligmer & PA Feature& State & $d_1 \rightarrow d_1^{*}$ &
  $d_{1} \rightarrow d_{2}^*$ &
$d_{1} \rightarrow d_{3}^*$&
$(d_1 \rightarrow d_{1}^{*})^2$ &
$(d_1 \rightarrow l^{*})^2$  \\
   \hline \hline
PPP4 & I   & $2A_g$   & 0.5546 & 0.0387 &---&0.1121&--- \\
 & II  & $3A_g$   & 0.0471 & 0.3396 &---&0.2164&---\\
     &     & $4A_g$   & 0.0779 & 0.1787&---&0.2997 &0.0681 \\
     & III & $5A_g$   & 0.0385 & 0.2474 &---&0.1037 & 0.0210 \\
     & IV  & $8A_g$   & ---    & 0.3476 &---&0.1566 &0.0147 \\
 
     &     & $10A_g$  &  ---   & 0.1848 &---& 0.0894 & 0.3289 \\\hline 
PPP5 & I   & $2A_g$   & 0.6171 & ---    &---&0.0903 
& --- \\
 
   & II & $4A_g$ & 0.1236 & ---&---&0.4688 &0.0695 \\
     & III & $7A_g$ & 0.1309 & --- &---&0.4390 &  --- \\
 
    & IV  & $9A_g$   & ---    & 0.2284 & --- &  0.0543 
& 0.2516 \\
\hline
PPV3 & I   & $2A_g$& 0.3872  &---&---&0.2847 &---  \\
& II  & $4A_g$&0.3042 &0.0488&---&0.3313   &0.0169 \\
    & III & $7A_g$ &0.2732&0.1458&--- &--- & --- \\
& IV&10$A_g$&---&0.0338&0.1746&0.0256 &0.3193 \\ \hline
PPV4 & I&2$A_g$& 0.5078&---&---&0.1884& ---\\
&II&3$A_g$& 0.1983&---&---&0.4680 &---\\
&III& 4$A_g$ & 0.4178&---&---&0.2483 &---\\
&& 5$A_g$& 0.0149&0.3366&---&0.2859& ---\\
&IV&9$A_g$&0.0602&0.4555&---&0.0318& 0.0398\\      
&V&12$A_g$& ---&0.1140 &0.1460&0.02071& 0.2284\\ \hline  
\end{tabular}
\label{tab-coeff}
\end{table}
%
%
%\end{document}

%
\end{document}